# Pressure-induced Superconductivity in Noncentrosymmetric Weyl Semimetals LaAlX (X = Si, Ge)

Weizheng Cao[1#], Ningning Zhao[2#], Cuiying Pei[1], Qi Wang[1,3], Yi Zhao[1], Lingling Gao[1], Changhua Li[1], Na Yu[1], Yulin Chen[1,3,4], Kai Liu[2*], and Yanpeng Qi[1,3*]

1. School of Physical Science and Technology, ShanghaiTech University, Shanghai 201210, China
2. Department of Physics and Beijing Key Laboratory of Opto-electronic Functional Materials & Micro-nano Devices, Renmin University of China, Beijing 100872, China
3. ShanghaiTech Laboratory for Topological Physics, ShanghaiTech University, Shanghai 201210, China
4. Department of Physics, Clarendon Laboratory, University of Oxford, Parks Road, Oxford OX1 3PU, UK

\# These authors contributed to this work equally.
\* Correspondence should be addressed to Y.Q. (qiyp@shanghaitech.edu.cn) or K.L. (kliu@ruc.edu.cn)

## ABSTRACT

In topological materials, Dirac fermions can split into two Weyl fermions with opposite chiralities due to the breaking of space inversion symmetry, while in non-centrosymmetric superconductors, novel superconducting electron pairing mechanisms arise because of the antisymmetric spin-orbit coupling. In this work, we report the pressure-introduced superconductivity in a typical noncentrosymmetric Weyl semimetal LaAlX (X=Si and Ge). Superconductivity was observed at around 65 GPa without structural phase transition. A typical dome-shape phase diagram is obtained with the maximum $T_c$ of 2.5 K (2.1 K) for LaAlSi (LaAlGe). Furthermore, the application of pressure does not destroy the nontrivial band topology of LaAlSi up to 80.4 GPa, making such materials as potential candidates for realizing topological superconductivity. Our discovery of superconductivity in LaAlX (X=Si and Ge) will provide critical insight in noncentrosymmetric superconductors and stimulate further study on superconductivity in Weyl semimetals.

# INTRODUCTION

Superconductivity in systems lacking space inversion symmetry, so-called noncentrosymmetric superconductor (NCSs) [1], has been extensively studied since the discovery of the coexistence of heavy fermion superconductivity and antiferromagnetism in CePt$_3$Si [2-4]. The lack of space inversion symmetry in NCSs together with the corresponding antisymmetric spin-orbit coupling (ASOC) removes the parity constraint on the Cooper pair and allows for a mixture of spin-singlet and spin-triplet states [5]. This unusual Cooper pair formation results in an unconventional superconductivity which may manifest itself by the gap with nodal lines or points, the multiband effects, or the extraordinarily high critical magnetic field [6-8].

Up to now, a number of NCSs have been found. Among them, ternary equiatomic phosphides *MM'P* compose a large compound family and are noted for their relatively high $T_c$ values [9-12]. Here *M* can be an early transition metal element or a rare earth element, *M'* is a late transition metal element, and *P* is a pnictogen. Different from most *MM'P* NCSs with the ZrNiAl-type structure, we reported superconductivity first time in ternary equiatomic pnictides La*M'P* (*M'* = Ir and Rh; *P* = P and As) that crystallize in the tetragonal LaPtSi-type structure[13]. Later, actinide ThIrP with the same structure was also synthesized at ambient pressure and shows superconductivity at low temperature[14]. These results indicate that the compounds with the LaPtSi-type structure are potential candidates to explore NCSs. Recently, rare-earth based compounds RAlX (R=La, Ce, Pr, Nd, and Sm; X=Si, Ge) [15-27] with the LaPtSi-type structure have attracted widespread interest. With the naturally broken space-inversion symmetry, the noncentrosymmetric RAlX compounds have been theoretically predicted and experimentally confirmed as a new family of Weyl semimetals [17, 22, 28]. Realization of superconductivity in those noncentrosymmetric topological materials is particularly intriguing since they provide a natural platform to realize Majorana fermions [29, 30], the manipulation of which forms the basis of future topological quantum computing [31].

In general, the simultaneous realization of nontrivial band topology and superconductivity in the same material under ambient condition remains rare. Another feasible route to find so-called topological superconductor is to search for intrinsic superconductivity in topological materials by chemical doping or high pressure. In comparison with chemical doping, high pressure is a "clean" and powerful tool to effectively modify electronic states in a systematic fashion. Indeed, superconductivity achieved in this manner has been observed in some topological compounds [32-37]. Here, we systematically investigate both the structures and the electronic properties of noncentrosymmetric Weyl semimetals LaAlX (X = Si, Ge) under high pressures up to 95.1 GPa. The crystal structure of LaAlSi without space-inversion symmetry is robust and persists up to 80.4 GPa. Interestingly, we observed the appearance of superconductivity in LaAlX (X = Si, Ge) at around 65 GPa without structural phase transition. A typical dome-sharp phase diagram is obtained with the maximum $T_c$ of 2.5 K and 2.1 K for LaAlSi and LaAlGe, respectively. The application of pressure does not destroy the nontrivial band topology of LaAlSi up to 80.4 GPa. The possibility of topological superconductivity in LaAlX (X = Si, Ge) is also discussed.

# EXPERIMENTAL DETAILS

The single crystals of LaAlX (X = Si, Ge) were grown via self-flux method [15, 21]. High-purity block of lanthanum, aluminum and silicon (or germanium), were mixed in a molar ratio of 1 : 10 : 1 and loaded into an alumina crucible. All treatments were performed in an argon-filled glove box, where both $H_2O$ and $O_2$ were limited to less than 0.5 ppm, and then the crucible was sealed in a quartz tube under vacuum. The quartz tube was heated up to 1100 °C in 24 h with temperature holding for 12 h in order to ensure guarantee of the raw material melting. Subsequently, the temperature slowly cooled down to 750 °C at a rate of 2 °C/h and then excess Al was removed by a high-speed centrifuge. Large plates of LaAlX (X = Si, Ge) single crystals were obtained with a typical dimension of $5 \times 5 \times 1$ mm$^3$.

The phase identification of the resulting samples was carried out by powder x-ray diffraction (PXRD) with Cu Kα at room temperature. Rietveld refinements of the PXRD patterns were performed using the TOPAS code[38]. The crystal surface morphology and composition were examined by scanning electron microscopy (SEM) and energy dispersive X-ray (EDX) analysis. Electrical transport properties including resistance, magnetoresistance, and Hall effect were performed on Physical Property Measurement System (PPMS). High-pressure electrical transport measurements were performed in a nonmagnetic diamond anvil cell (DAC). Four platinum sheet electrodes were touched to the sample for resistance measurements with van der Pauw method. Pressure was determined by the ruby luminescence method [39]. High-pressure XRD diffraction experiments were performed at beamline BL15U of Shanghai Synchrotron Radiation Facility (X-ray wavelength λ = 0.6199 Å). Mineral oil was used as pressure transmitting medium. The two dimensional diffraction images were analyzed using the FIT2D software [40]. Rietveld refinement of crystal structures under high pressure using the General Structure Analysis System (GSAS) and the graphical user interface (EXPGUI) [41, 42].

The first-principles electronic structure calculations on LaAlSi were performed by using the projector augmented wave (PAW) method [43, 44] as implemented in the VASP package [45, 46]. The generalized gradient approximation (GGA) of Perdew-Burke-Ernzerhof (PBE) type [47] was used for the exchange-correlation functional. The kinetic energy cutoff of the plane-wave basis was set to be 320 eV. The Brillouin zone was sampled with an 8×8×8 *k*-point mesh. The Gaussian smearing method with a width of 0.05 eV was adopted for the Fermi surface broadening. The lattice constants of LaAlSi under high pressure were fixed to the experimental values. The spin-orbit coupling (SOC) effect was included in the band structure calculations. The topological charge (Chern number) for the Weyl points was computed by the Wilson-loop technique [48, 49] with the WannierTools package [50]. The tight-binding Hamiltonian was constructed by the maximally localized Wannier functions [51, 52] for the outmost *s*, *p*, *d* orbitals of La atoms, *s*, *p* orbitals of Al atoms, and *s*, *p* orbitals of Si atoms generated by the first-principles calculations.

**RESULTS AND DISCUSSION**

Prior to physical properties measurements, the synthesized LaAlX (X = Si, Ge) samples were structurally characterized by using powder x-ray diffraction (PXRD). As shown in Figs. 1(a) and 1(b), all the Bragg peaks can be indexed into tetragonal structure with the space group $I4_1md$ (No. 109). The inset in Fig. 1(a) shows the crystal structure of LaAlX (X = Si, Ge), which is an ordered ternary derivative structure of the α-ThSi$_2$ type [53-55]. The Al and Si/Ge atoms are linked to form a three-dimensional network with a trigonal planar environment, whereas the La atoms are located in the cavities. The sublattice of Al atoms does not possess an inversion center along the *c* axis; thus, the space inversion symmetry is broken. The average compositions were derived from a typical EDX measurement at several points on the crystal, revealing good stoichiometry with the atomic ratio of La : Al : Si = 29.65 : 35.86 : 34.49. The XRD refinement supports the noncentrosymmetric structure. The refined lattice parameters of LaAlX (X = Si, Ge) are summarized in Table I, which are in good agreement with the previously reported data [15, 18].

Both LaAlSi and LaAlGe show typical metallic behaviors. The temperature dependences of longitudinal resistivity ($\rho_{xx}$) and Hall resistivity ($\rho_{yx}$) of LaAlSi are measured with the magnetic field along the [001] direction, as shown in Figs. 1(c) and 1(d), respectively. The positive magnetoresistance is observed in all the temperature range and reaches ~83% at 1.8 K and 9 T. The $\rho_{yx}$ exhibits a nearly linear dependence on magnetic field with a positive slope, indicating the dominant role of hole carriers for the transport [Fig. 1(e)]. By a linear fit to the Hall resistivity with the single band model, i.e., $\rho_{yx} = R_H B$, where $R_H = 1/en$ is the Hall coefficient, *n* donates the carrier density, and *e* is the electron charge, the temperature dependence of carrier density *n* and carrier mobility *μ* are obtained [Fig. 1(f)], which is consistent with previous report [15]. A similar evolution of $\rho_{xx}$ and $\rho_{yx}$ in external magnetic fields is observed for LaAlGe (see Fig. S1). It should be noted that with the increasing temperature the Hall coefficient of LaAlSi decreases, while that of LaAlGe increases.

Since no superconductivity was observed in LaAlX (X = Si, Ge) down to 1.8 K at ambient pressure, a question arises naturally: is it possible to achieve superconductivity in LaAlX (X = Si, Ge) under high pressure? Hence, we measured electrical resistivity $\rho(T)$ of LaAlX (X = Si, Ge) single crystals at various pressures. Figure 2(a) shows the typical $\rho(T)$ curves of LaAlSi for pressure up to 95.1 GPa. The $\rho(T)$ curves display a metallic-like behavior in the whole pressure range. Increasing pressure induces a continuous suppression of the overall magnitude of $\rho$. When the pressure increases to around 62.1 GPa, a small drop of $\rho$ is observed at the lowest measuring temperature ($T_{min}$ = 1.8 K), as shown in Fig. 2(b). With further increasing pressure, zero resistivity is achieved at low temperature for *P* > 73.4 GPa, indicating the emergence of superconductivity. The superconducting $T_c$ increases with pressure and a maximum $T_c$ of 2.5 K is attained at *P* = 78.5 GPa. Beyond this pressure $T_c$ decreases slowly, but superconductivity persists up to the highest experimental pressure of ~ 95.1 GPa. Interestingly, an abnormally increased $T_c$ as high as 2.8 K is observed in the decompression process and the superconducting transition remains until recovered to 19.5 GPa [Fig. S2(a)-(d)]. The fact that superconductivity retains after pressure released may suggest that LaAlSi transforms to a metastable phase, which shows superconductivity under high pressure and it keeps to low pressure [56, 57].

A similar evolution of $\rho(T)$ is observed for LaAlGe, and a maximum $T_c$ of 2.1 K is attained at *P* = 75.7 GPa [Figs. 2(c) and S3]. To gain insights into the superconducting transition, we measured $\rho(T)$ curves at various magnetic fields for LaAlX (X = Si, Ge).

Figure 2(d) demonstrates that the resistivity drop is continuously suppressed with increasing magnetic field and the superconducting transition could not be observed above 1.8 K at around 0.5 T. Such behavior further confirms that the sharp decrease of $\rho(T)$ should originate from a superconducting transition. The upper critical field, $\mu_0 H_{c2}$, is determined using 90% point on the resistivity transition curves, and plots of $H_{c2}(T)$ are shown in Fig. 2(f). A simple estimate using the conventional one-band Werthamer-Helfand-Hohenberg approximation, neglecting the Pauli spin-paramagnetism effect and spin-orbit interaction[58], yields a value of 1.03 T for LaAlSi. By using the Ginzburg–Landau (G-L) formula $\mu_0 H_{c2}(T) = \mu_0 H_{c2}(0) (1 − (T/T_c)^2)/(1 + (T/T_c)^2)$ to fit the data, the estimated $\mu_0 H_{c2}(0)$ value is 1.24 T at 78.5 GPa. A similar upper critical field is obtained in LaAlGe [Fig. 2(e)]. Although the crystal structure of LaAlX (X = Si, Ge) breaks space inversion symmetry, their upper critical fields are much lower than the Pauli limiting fields, $H_P(0) = 1.84 T_c$, respectively, indicating that Pauli pair breaking is not relevant.

At ambient pressure, LaAlX (X = Si, Ge) crystallize a noncentrosymmetric LaPtSi-type structure. To further identify the structural stability, we chose LaAlSi to perform *in situ* high-pressure XRD measurements. Figure 3(a) displays the high-pressure synchrotron XRD patterns of LaAlSi measured at room temperature up to 91.9 GPa. A representative refinement at 22.3 GPa is displayed in Fig. 3(b). All the Bragg reflections can be nicely refined by using the space group $I4_1md$ (No.109) as the initial model. As shown in Fig. 3(c), both *a*-axis and *c*-axis lattice constants decrease with increasing pressure. A third-order Birch-Murnaghan equation of state was used to fit the measured pressure-volume (*P-V*) data for LaAlSi (Fig. S4) [59]. The obtained bulk modulus $K_0$ is 105.85 GPa with $V_0$ = 272.95 Å$^3$ and $K_0'$ = 3.61. The structure of LaAlSi is robust and there is no structural phase transition until 80.4 GPa. It should be noted that the superconductivity is observed lower than this pressure. However, when the pressure increases up to 86.8 GPa, a new broad peak appears at ≈ 15.7° in the diffraction patterns. This indicates that the applied high pressures introduced amorphous phases into the material. Similar phenomenon was observed in some other topological materials [60-62]. The data are also collected when the sample is decompressed to ambient pressure [Fig. S2(e)]. The crystallinity of the released sample was confirmed by transmission electron microscope [Fig. S2(f)]. After a full pressure release, high pressure amorphous phase quenched to ambient pressure and did not recover to the LaPtSi-type structure.

At ambient pressure, LaAlX (X = Si, Ge) have been proposed as a new family of Weyl semimetal. To theoretically understand the evolution of topological properties under pressures, we have performed density functional theory (DFT) calculations on LaAlSi in the LaPtSi-type structure. Figures 4 and S5 show the electronic band structures and the Weyl points in the Brillouin zone of LaAlSi under various pressures. From the band structures calculated in the absence of SOC [Figs. 4(a), 4(d), 4(g)], we can see that there are two bands crossing the Fermi level and forming nodal lines around the Brillouin zone boundary, similar to the case at ambient pressure in a previous study. With the inclusion of SOC [Figs. 4(b), 4(e), 4(h)], the nodal lines are gapped and evolve into Weyl nodes carrying the Chern number *C* of ±1. According to our calculations, there are totally 22 pairs of Weyl points at 50.4 GPa [Fig. 4(c)] and 12 pairs of Weyl points at both 62.4 GPa and 73.6 GPa [Figs. 4(f) and 4(i)], which locate within $\pm 0.15$ eV near the Fermi level (Table SI). The Weyl semimetals with intrinsic superconductivity are predicted to be 3D topological superconductor candidates due to the nontrivial Fermi surface originating from the Weyl points [63, 64]. In consideration of the intrinsic superconductivity under pressures and the Weyl points near the Fermi level, LaAlX (X = Si, Ge) may offer a very promising material platform to realize the topological superconductivity.

On the basis of the above results, we can establish a $T$-$P$ phase diagram for LaAlSi single crystal. With the increasing pressure, the overall magnitude of resistivity is suppressed continuously. Superconductivity was observed at around 50 GPa without structural phase transition. The superconducting transition temperature $T_c$ increases with applied pressure and reaches a maximum value of 2.5 K at 78.5 GPa for LaAlSi (2.1 K at 75.7 GPa for LaAlGe), and a typical dome-like evolution of $T_c$ is obtained. The transport measurements on different samples for several independent runs provide the consistent and reproducible results (Fig. S6), confirming this intrinsic superconductivity under pressure. In addition, we did not observe other peaks from elements (eg, La or Al) or binary compounds [65, 66], the possibility of high-pressure decomposition of LaAlSi is hence ruled out. At ambient pressure, LaAlX (X = Si, Ge) is a Weyl semimetal in which there are many symmetry-related Weyl points nearest the Fermi level. The application of pressure does not destroy the nontrivial topology of LaAlSi up to 80.4 GPa. Above 86.8 GPa, a pressure-induced amorphization emerges. It is very interesting that an amorphous phase of LaAlSi could support superconductivity and retain after the pressure released. This will stimulate further studies from both experimental and theoretical perspectives.

**CONCLUSION**

In summary, superconductivity was successfully induced without structural phase transition in the noncentrosymmetric Weyl semimetals LaAlSi and LaAlGe by application of high pressure. The robustness of Weyl fermion is accompanied by the appearance of superconductivity, making LaAlX (X = Si, Ge) as possible candidates to realize topological superconductivity. Thus, the noncentrosymmetric crystal structure, the nontrivial topological state, and the superconductivity were all observed in LaAlX (X = Si, Ge), all contributing to the highly interesting physics in this rare-earth-based compounds RAlX.

**ACKNOWLEDGMENT**

This work was supported by the National Key R&D Program of China (Grants No. 2018YFA0704300 and 2017YFA0302903), the National Natural Science Foundation of China (Grant No. U1932217, 11974246, 12004252, 11774424, and 12174443), the Natural Science Foundation of Shanghai (Grant No. 19ZR1477300), the Science and Technology Commission of Shanghai Municipality (19JC1413900), the Fundamental Research Funds for the Central Universities (CN), and the Research Funds of Renmin University of China (Grant No. 19XNLG13). N.Z. is supported by the Outstanding Innovative Talents Cultivation Funded Programs 2022 of Renmin University of China. The authors thank the support from Analytical Instrumentation Center (# SPST-AIC10112914), SPST, ShanghaiTech University. The authors thank the staffs from BL15U1 at Shanghai Synchrotron Radiation Facility for assistance during data collection. Computational resources were provided by the Physical Laboratory of High Performance Computing at Renmin University of China.

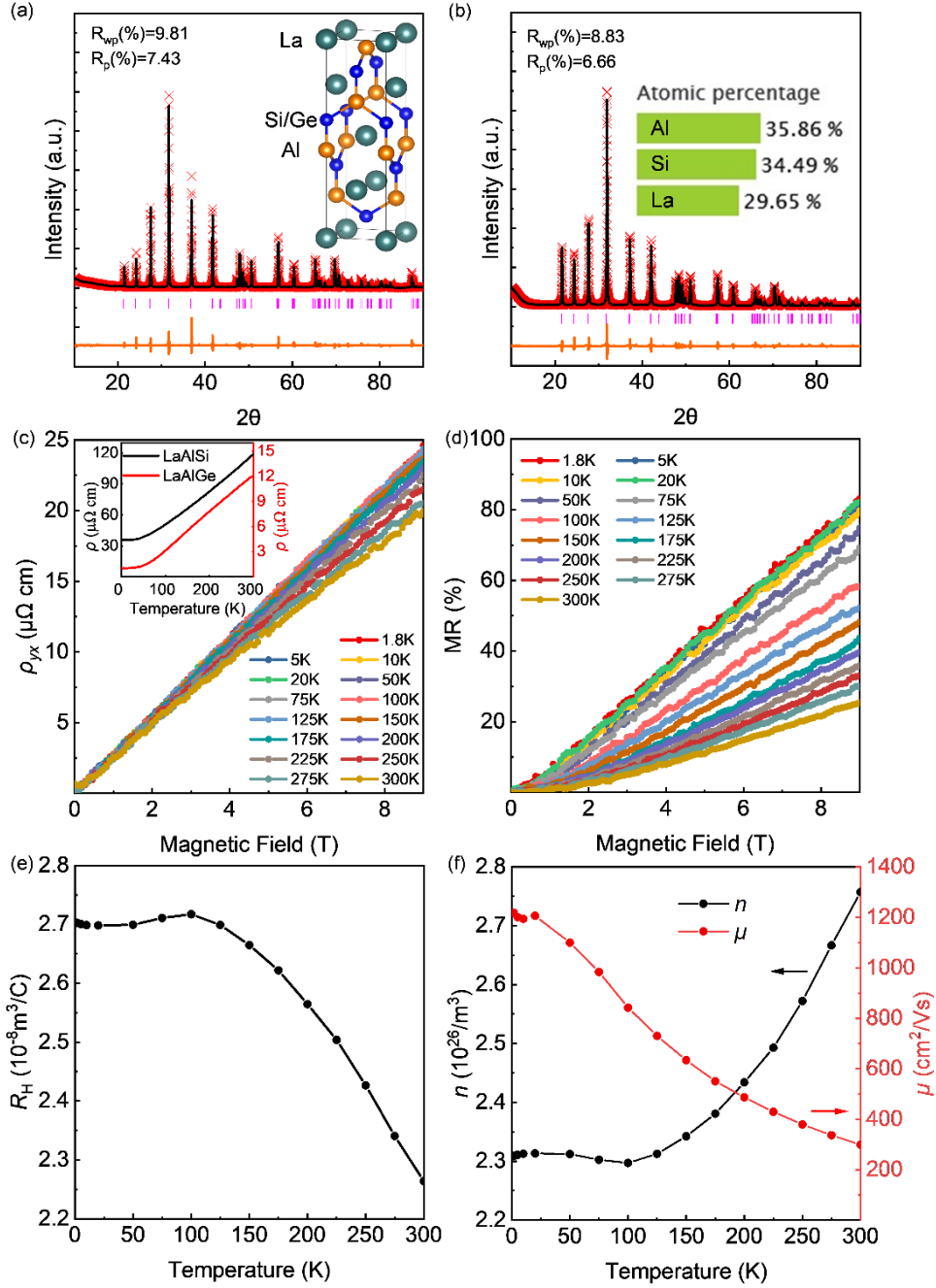

FIG. 1. (a) and (b) Powder XRD pattern of LaAlGe and LaAlSi at room temperature. Insets: the structure of LaAlX (X=Si and Ge) and the elemental content of LaAlSi. (c) Hall resistivity and (d) Longitudinal resistivity as a function of magnetic field B at various temperature for LaAlSi. The magnetoresistance (MR) is defined as MR = $[\rho(B) - \rho(0)]/\rho(0) \times 100\%$ in which $\rho(B)$ and $\rho(0)$ represent the resistivity with and without B, respectively. Inset of (c): resistivity dependence of temperature for LaAlX. (e) The temperature dependent Hall coefficient for LaAlSi. (f) Carrier concentration and mobility of LaAlSi as a function of temperature, respectively.

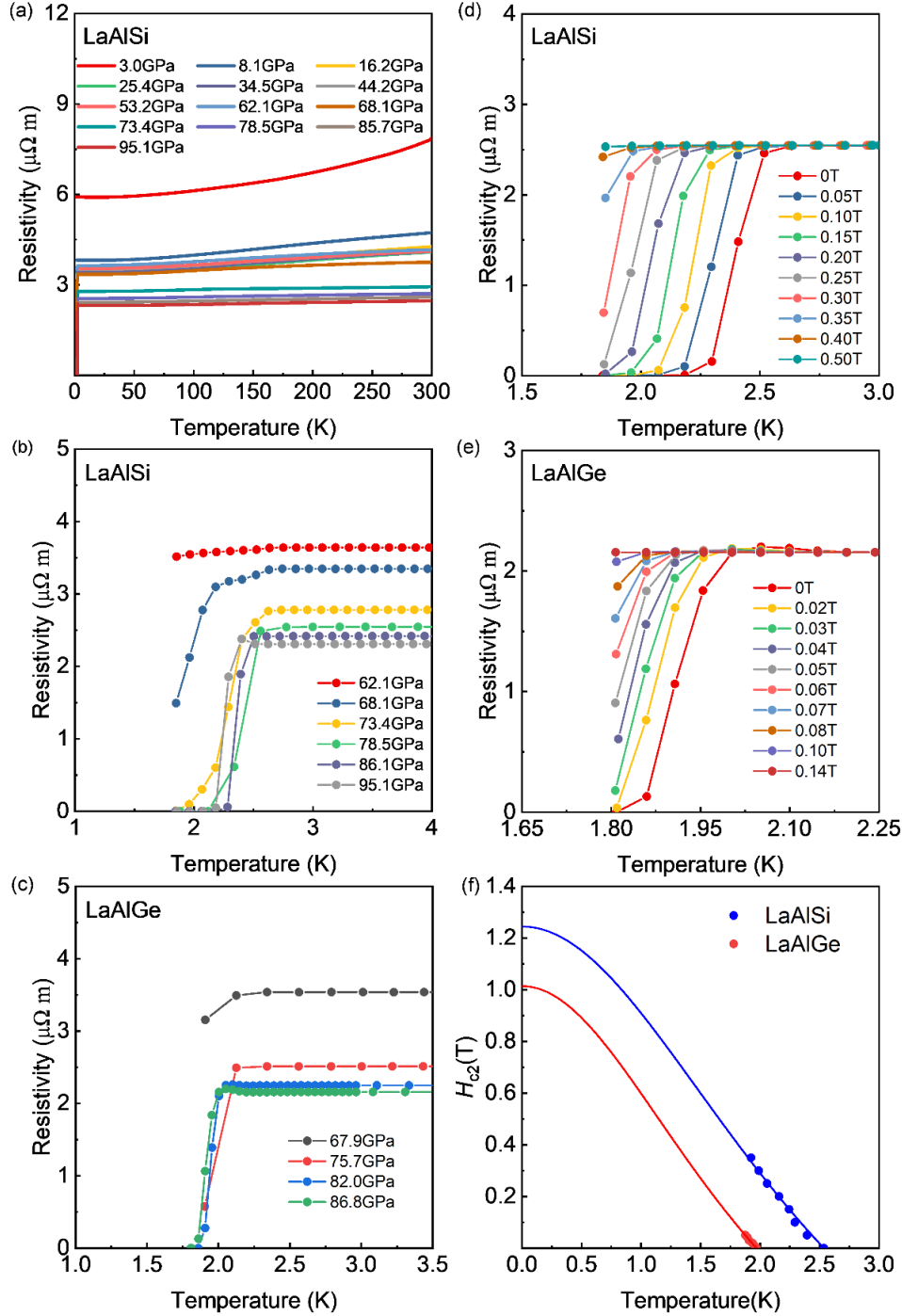

FIG. 2. (a) Pressure dependence of electrical resistivity of LaAlSi at temperature range of 1.8 - 300 K in run 4. (b) Pressure-induced superconductivity. The temperature-dependent resistivity under various pressures from 62.1 GPa to 95.1 GPa. (c) Temperature dependence of resistivity measured at various pressure between 67.9 GPa and 86.8 GPa for LaAlGe. (d) Resistivity as a function of temperature at the pressure of 78.5 GPa under different magnetic fields for LaAlSi in run 4. (e) Temperature dependence of resistivity for various magnetic fields for LaAlGe at 86.8 GPa. (f) $H_{c2}$ as a function of temperature at the pressure of LaAlSi at 78.5 GPa and LaAlGe at 86.8 GPa, respectively. $T_c$ is defined as the critical transition temperature at which resistivity reduced to 90% of its normal value. The curve represents the Ginzburg-Landau (GL) fits.

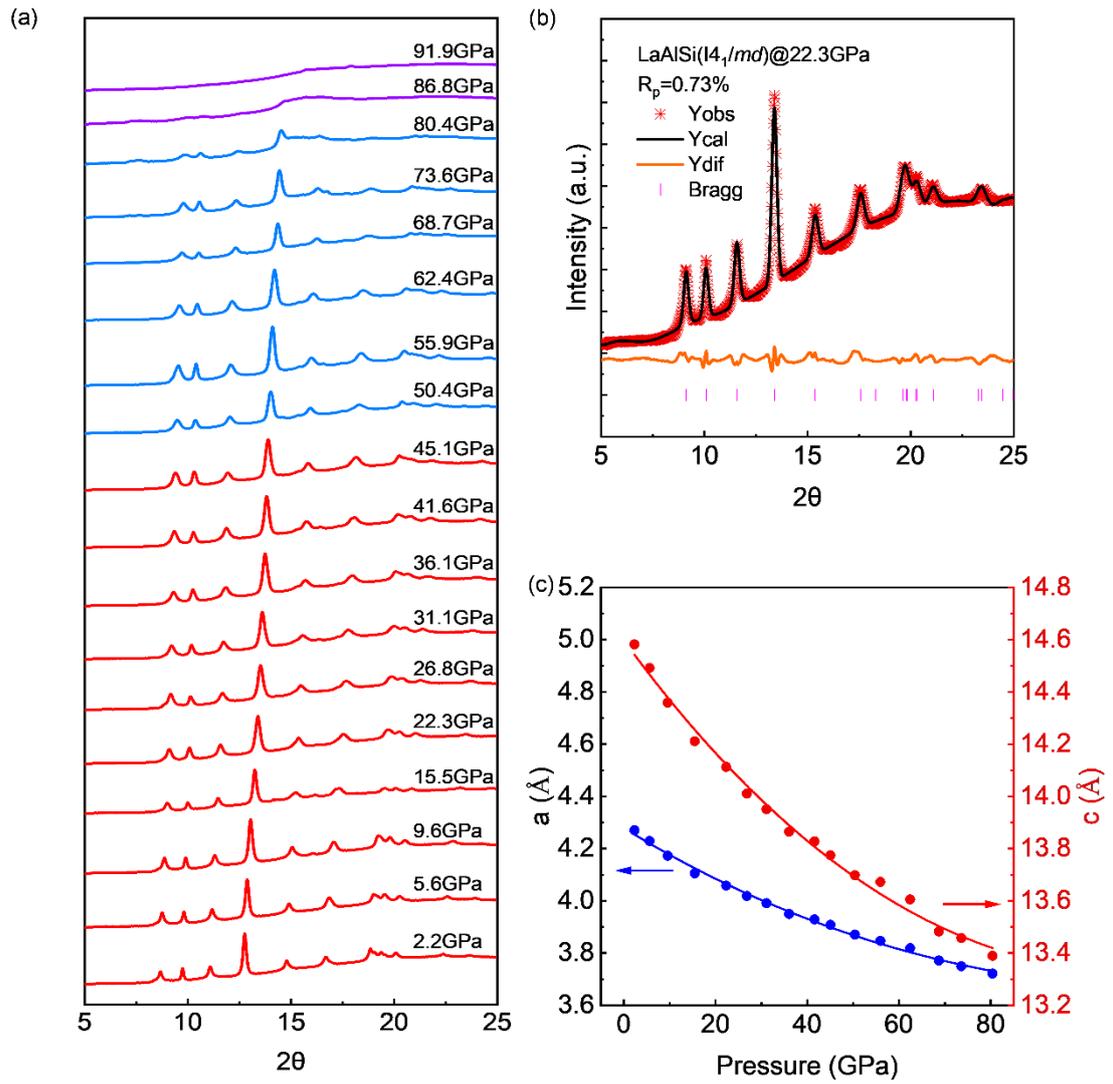

FIG. 3. (a) XRD patterns of LaAlSi under different pressures up to 91.9 GPa. The structure of LaAlSi is amorphous above 86.8 GPa. (b) High-pressure x-ray diffraction of LaAlSi and Rietveld refinement. (c) Pressure dependence of the lattice constants $a$ and $c$ for LaAlSi.

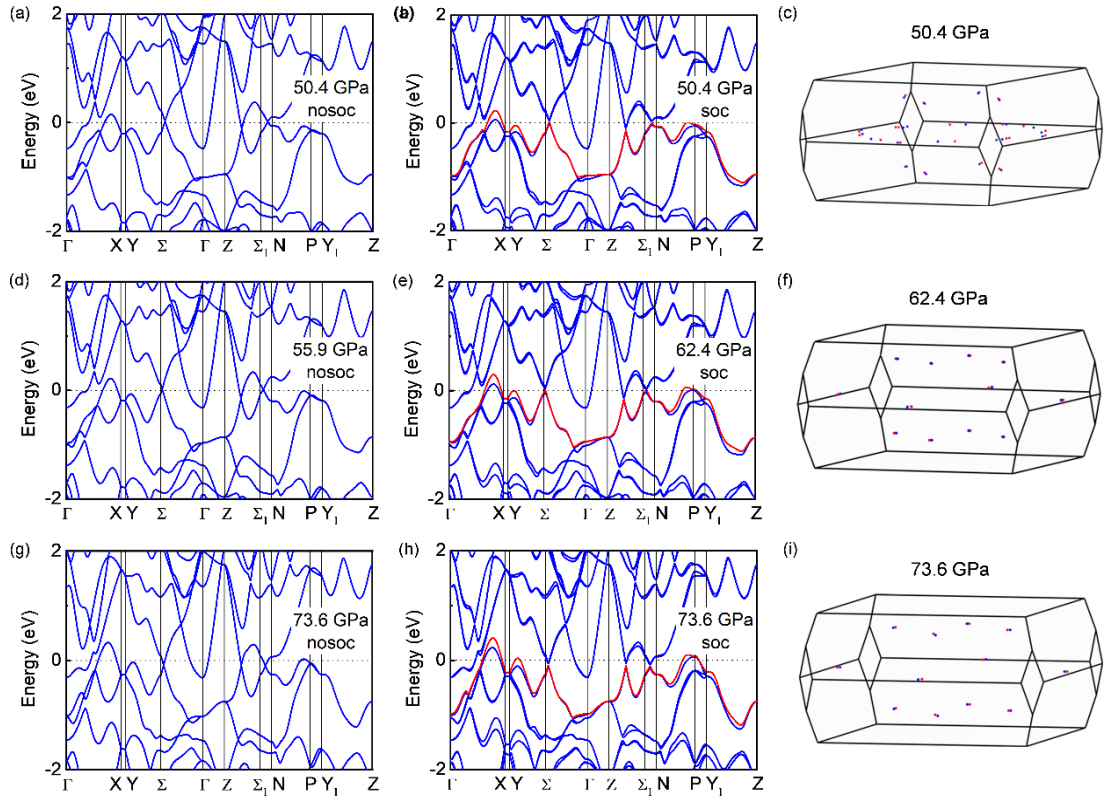

FIG. 4. Electronic band structures of LaAlSi calculated without and with the inclusion of the SOC effect at (a)(b) 50.4 GPa, (d)(e) 62.4 GPa, and (g)(h) 73.6 GPa, respectively. The red lines represent the highest valence bands. (c)(f)(i) The Weyl points in the Brillouin zone which locate within ±0.15 eV near the Fermi level under 50.4 GPa, 62.4 GPa, and 73.6 GPa, respectively. The red and blue dots represent the Weyl points with chiralities of +1 and -1, respectively.

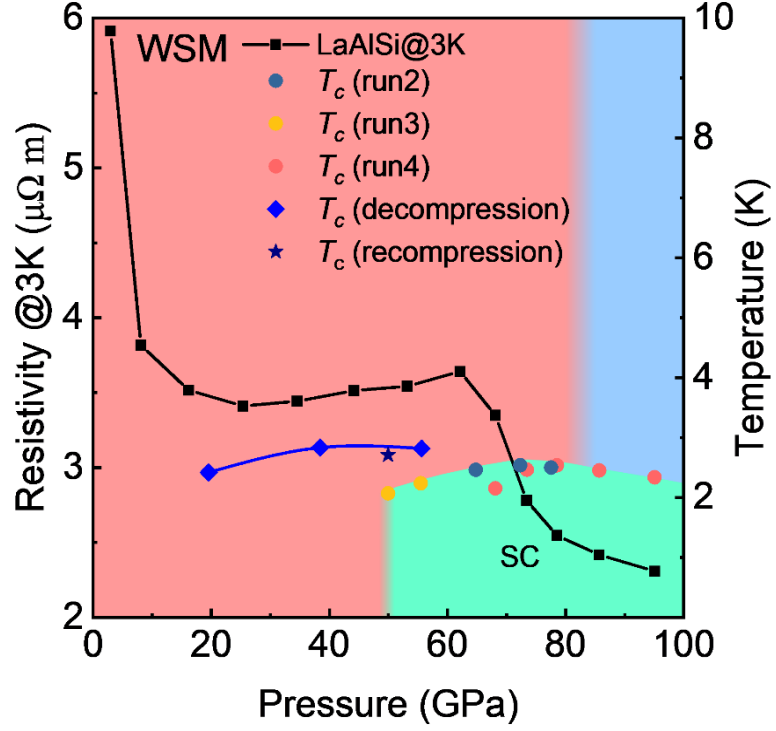

FIG. 5. Phase diagram depicting pressure dependence of resistivity at 3 K for LaAlSi and the $T_c$ from different resistivity measurements for LaAlSi. A dome-like evolution is observed with a maximum $T_c$ of 2.5 K at 78.5 GPa for LaAlSi. The resistivity at 3 K for LaAlSi displays a more complicated feature. The WSM state persists until 80 GPa after that the crystal structure becomes amorphous.

TABLE I. Structure parameters and superconductivity of LaAlX (X=Si, Ge)

| Compound | LaAlSi | LaAlGe |
| --- | --- | --- |
| Space group | $I4_1md$ (No. 109) | |
| Crystal structure | LaPtSi-type | |
| Lattice parameter $a$ | 4.3124 Å | 4.3938 Å |
| Lattice parameter $c$ | 14.6772 Å | 14.8285 Å |
| $T_{c\,max}$ | 2.5 K | 2.1 K |
| $H_{c2}$(WHH) | 1.03 T | 0.71 T |
| $H_{c2}$(GL) | 1.24 T | 1.01 T |
| $\xi_{GL}(0)$ | 16.27 nm | 18.03 nm |